\documentstyle[eqsecnum,aps]{revtex}
\input epsf.sty

\makeatletter
\@addtoreset{equation}{section}
\makeatother

\newcommand{\bibi}{\bibitem}

\def\g{\gamma}

\def\j{\psi}
\def\k{\kappa}     

\def\m{\mu}

\def\n{\nu}

\def\p{\pi}       

\def\x{\xi}

\def\S{\Sigma}

\def\jb{\overline{\j}}

\newcommand{\half}{\mbox{{\normalsize $\frac{1}{2}$}} }

\newcommand{\quart}{\mbox{{\small $\frac{1}{4}$}} }

\newcommand{\ra}{\rightarrow}

\newcommand{\lag}{\langle}
\newcommand{\rag}{\rangle}

\newcommand{\tk}{\widetilde{\kappa}}

\newcommand{\Dsl}{D\!\!\!\!/}

\newcommand{\dsl}{\partial \!\!\!/}
\newcommand{\pslash}{p \!\!\!/}

\newcommand{\jn}{\j^{{\rm n}}}
\newcommand{\jbn}{\jb^{{\rm n}}}
\newcommand{\jc}{\j^{{\rm c}}}
\newcommand{\jbc}{\jb^{{\rm c}}}

\newcommand{\hmu}{\hat{\mu}}

\newcommand{\be}{\begin{equation}}
\newcommand{\ee}{\end{equation}}
\newcommand{\bea}{\begin{eqnarray}}
\newcommand{\eea}{\end{eqnarray}}
\newcommand{\eq}{\ref}
\newcommand{\beq}{\begin{equation}}
\newcommand{\eeq}{\end{equation}}

\newcommand{\lb}{\label}

\def \3{\ss}

\begin{document}
\draft
\preprint{TAUP-2454-97, HU-EP-97/60, Wash.~U.~HEP/97-62} 
\twocolumn[\hsize\textwidth\columnwidth\hsize\csname%
@twocolumnfalse\endcsname
\title{Lattice Chiral Fermions Through Gauge Fixing  
}
\author{Wolfgang Bock}
\address{
Institute of Physics, Humboldt University Berlin, \\
Invalidenstr. 110, 10115 Berlin, Germany 
}
\author{Maarten F.L. Golterman}
\address{
Department of Physics, Washington University, \\
St. Louis, MO 63130, USA
}
\author{Yigal Shamir}
\address{
School of Physics and Astronomy, Beverly and Raymond Sackler 
Faculty \\
of Exact Sciences Tel-Aviv University, Ramat Aviv 69978, Israel
}
\date{\today}
\maketitle
\begin{abstract}
We study a concrete lattice regularization of a U(1) chiral gauge theory. 
We use Wilson fermions, and include 
a Lorentz gauge-fixing term and a gauge-boson mass counterterm.
For a reduced version of the model, in which the gauge fields are constrained 
to the trivial orbit, we show that there are no species doublers, and 
that the fermion spectrum contains only the desired states 
in the continuum limit, namely  charged left-handed (LH) fermions
and neutral right-handed (RH) fermions.
\end{abstract}
]                                                    
\narrowtext
\section{Introduction}
\label{SEC1}
While vectorlike theories like QCD can be formulated on the lattice in a
manifestly gauge invariant way, this is not the case for chiral gauge
theories.  The reason for this is the anomaly, which forces the
lattice regularization of a chiral gauge theory to break chiral
invariance, so that the contribution to the anomaly for each fermion 
species is recovered for smooth gauge fields \cite{KaSm81}.  
For generic
lattice gauge fields, though, the breaking is more severe, and it turns out
that  the coupling of the lattice gauge degrees of freedom
to the fermions destroys their chiral nature.  

This ``problem of
rough gauge fields" has been the central obstruction to the construction
of lattice chiral gauge theories to date \cite{Sh96r}.  It is therefore natural
to try control the effects of the gauge degrees of freedom by fixing the
gauge.  It has been proposed to extend the usual
perturbative definition of chiral
gauge theories, based on gauge fixing and 
BRST invariance, to the lattice
\cite{BoMa89}.  On the lattice, the fermion action 
breaks BRST invariance explicitly.  The hope is then
that a gauge-invariant continuum limit (CL) can be defined by suitable
adjustment of the coefficients of a finite set of counterterms.

The lattice gauge-fixing
action must have a unique absolute minimum 
when the compact lattice link variables are set to $U_{\mu x}=I$, in order
to be able to use weak-coupling perturbation theory.
Such lattice discretizations were given first for a nonlinear gauge \cite{Sh95}
and later for the Lorentz gauge \cite{GoSh96}.
It was argued that (in both
cases) a continuous phase transition will occur between two different
phases with broken symmetry. The gauge symmetry, however, 
should be restored {\it at} 
the transition, and it is here that the fermion spectrum is expected to be
chiral.  For a U(1) theory (where ghosts are not needed), the
existence of this phase transition was demonstrated in ref. \cite{BoGoSh97}.
In this Letter, we address the fermion spectrum of the U(1) case.
(For recent work in two dimensions using a different 
(``interpolation") approach, 
see ref. \cite{BoHe97}.)
\section{The Model}
\label{SEC2}
The U(1) lattice model we will consider is defined by the path integral 
\bea
&& Z=\int D U D \jb D \j \; e^{-S(U;\jb,\j)} \;, \lb{PATH} \\
&& S= \sum_x \{ {\cal L}_{\rm g}(x)+{\cal L}_{\rm f}(x)
           +{\cal L}_{{\rm g.f.}}(x)
                   +{\cal L}_{{\rm c.t.}}(x) \}\;, \lb{ACTION} \\
&& {\cal L}_{\rm g} = \frac{1}{g^2} \; \sum_{\m \n} 
\left\{ 1-\mbox{Re } U_{\m \n x} \right\} \;,  \lb{SG} \\
&& {\cal L}_{\rm f} =   \jb \{ \Dsl\,(U) P_{\rm L} + \dsl P_{\rm R} \} \j
-\frac{r}{2}\; \jb \Box \j  \;, \lb{SF} \\
&& {\cal L}_{\rm g.f.} =  \tk \{ \;              
\sum_{y} \; [\Box^2(U)]_{xy}  - B_x^2(U)  \; \} \; , \; \;   \tk=\frac{1}{2 \x 
g^2 } 
\;, \lb{SGF} \\
&& {\cal L}_{{\rm c.t.}} = -\k  \sum_{\m} \left\{ U_{\m x}+ U_{\m x}^{\dagger}
  \right\} \;, \lb{SC} \\
%
%
%
%
&&B_x(U)=\quart \; \sum_\m ( V_{\m x-\hmu} + V_{\m x} )^2 \;, \;\;
V_{\m x} =  \mbox{Im}  \; U_{\m x}  \;. \lb{BX} 
\eea
$U_{\m x}=\exp(i a g A_{\m x})$ is the lattice 
link variable (we take the lattice spacing $a=1$), 
$U_{\m \n x}$ is the plaquette variable, 
$g$ is the gauge coupling, $r$ is the Wilson parameter, 
$\x$ is the gauge-fixing parameter and $P_{\rm L,R}=\half (1 \mp \g_5)$.    
$\partial_\m$ and $D_\m(U)$ designate the free and covariant antihermitian 
nearest-neighbor lattice derivatives, 
and $\Box$ and $\Box(U)$ the free and covariant 
nearest-neighbor lattice laplacians.     
We have added a Wilson term to the naive 
fermion lagrangian to remove the 15 unwanted 
species doublers situated 
at the corners of the four-dimensional Brillouin zone. 
Eq.~(\eq{SGF}) is the Lorentz gauge-fixing term 
introduced in ref.~\cite{GoSh96}.
We have included a gauge-boson mass 
counterterm, which is the only counterterm of dimension two, and 
ignored all dimension four counterterms which we believe to be  less 
important \cite{GoSh96}. It will become clear that 
a fermion-mass counterterm is not needed.

We now derive an equivalent form of the path integral. 
Using the invariance of the measure,
we first perform a gauge rotation, 
$\j_{{\rm L} x} \ra \phi_x^{\dagger} \j_{{\rm L} x}$, 
$U_{\m x} \ra \phi_x^{\dagger} U_{\m x} \phi_{x+\hmu}$. 
The Wilson term, gauge-fixing 
term and mass counterterm are 
not gauge invariant and therefore pick up factors of $\phi$.     
After integrating  over $\phi$ (using $\int d\phi_x=1$) we find 
\be
Z=\int D U D \phi D \jb D \j \; e^{-S(U;\phi;\jb,\j)} \;, \lb{PATH_PHI} \\
\ee
where $S(U;\phi;\jb,\j)$ is again given by eq.~(\eq{ACTION}) now with 
\bea
&& {\cal L}_{\rm f} = \jb \{ \Dsl P_{\rm L} + \dsl P_{\rm R} \} \j
-\frac{r}{2} \;   
\left\{ \jb \phi \Box P_{\rm R} \j  +{\rm h.c.} \right\} \lb{TSF} \;, \\
&& {\cal L}_{\rm g.f.} = \tk \left\{               
\phi^{\dagger}_x \sum_y [\Box^2(U)]_{xy} \phi_y - B_x^2(U;\phi) \right\} 
\; ,\lb{TSGF} \\
&& {\cal L}_{{\rm c.t.}} = -\k  \sum_{\m} 
\left\{ \phi_x^{\dagger} U_{\m x} \phi_{x+\hmu} + {\rm h.c.} \right\} 
\;. \lb{TSC}
\eea
$B_x(U;\phi)$ is given by eq.~(\eq{BX}) with 
$U_{\m x} \ra \phi_x^{\dagger} U_{\m x} \phi_{x+\hmu}$.
The plaquette term ${\cal L}_g$ and the fermion kinetic term
remain unchanged since they are gauge invariant. 
 The longitudinal gauge degrees of freedom are now associated
with the group-valued field $\phi$ 
which couples to the fermions through the Wilson term.
 Eq.~(\eq{PATH_PHI}) is invariant under 
the U(1)$_{\rm L}^{\rm local} \otimes$U(1)$_{\rm R}^{\rm global}$
symmetry $\j_{\rm L} \ra g_{{\rm L} x} \j_{{\rm L} x}$, 
          $\j_{\rm R} \ra g_{\rm R} \j_{{\rm R} x}$, 
          $U_{\m x}  \ra g_{{\rm L} x} U_{\m x} g_{{\rm L} x+\hmu}^{\dagger}$, 
          $\phi_{x}  \ra g_{{\rm L} x} \phi_x g_{\rm R}^{\dagger}$.
The model is 
also invariant under a shift-symmetry 
$\j_{\rm R}\rightarrow \j_{\rm R}+\epsilon_{\rm R}$, 
which implies that the fermion is massless,
and that its RH part decouples in the CL \cite{GoPe89}. 

The model without the gauge-fixing term ($\tk=0$)
is the Smit-Swift model. 
Its failure  to produce a chiral theory was demonstrated in a 
{\em reduced} version \cite{Sh96r}, in which $U_{\mu x}$ 
is set equal to one in eq. (\eq{PATH_PHI}), taking
only the dynamics of the gauge degrees of freedom $\phi$ 
into account. At large $r$, a phase exists
with unbroken U(1)$_{\rm L} \otimes$U(1)$_{\rm R}$ 
and without 
doublers. The spectrum in this phase, however, contains only
a {\it neutral} (under U(1)$_{\rm L}$) Dirac fermion 
$\jn= \j_{\rm R} + \phi^{\dagger} \j_{\rm L}$,
which will not couple to the gauge fields.
\section{Fermion Spectrum}
\label{SEC3}
In this Letter we will 
address the following important question: What does the fermion 
spectrum look like when we include the gauge-fixing term?
To obtain massless photons in the CL, 
the coefficient $\k$ of the gauge-boson mass counterterm in eq.~(\eq{SC}) 
has to be tuned to a critical  value $\k_c(\tk)$.  
Henceforth we will consider again the reduced version of the full model 
obtained by setting $U_{x \m}=1$ in eq.~(\eq{PATH_PHI}). 
In the reduced model, $\k=\k_c(\tk)$ is 
a continuous phase transition line separating
a {\em ferromagnetic} (FM) phase at $\k > \k_{{\rm c}}$,
and a so-called {\em ferromagnetic 
directional} (FMD) phase at $\k < \k_{{\rm c}}$ \cite{Sh95}, 
where  rotational symmetry is broken 
by a vector condensate, $\lag V_{\m x} \rag \ne 0$, cf.~eq.~(\ref{BX}). 
 This critical line exists in the interval $\tk>\tk_{{\rm TP}}>0$
and ends at a tricritical point located at $\tk= \tk_{{\rm TP}}$.      
To 1-loop order we find 
$\k_{{\rm c}} = 0.02993 + O(1/\tk)$ \cite{BoGoSh97}.
The U(1)$_{\rm L} \otimes$U(1)$_{\rm R}$ 
symmetry is broken to its diagonal subgroup in the FM and FMD phases, 
but is restored on the FM-FMD phase transition line because of
infra-red effects associated with a $1/(p^2)^2$ propagator for
the $\phi$-field fluctuations \cite{BoGoSh97} (we showed this by numerically
computing the order parameter $\langle\phi\rangle$ very close to the phase 
transition, finding very good agreement with 1-loop perturbation theory,
in which $\langle\phi\rangle\sim |\kappa-\kappa_c|^{1/(64\pi^2\tk)}$ for
large $\tk$). This symmetry restoration 
is an  essential prerequisite for the construction of a chiral gauge 
theory with {\it unbroken} gauge symmetry in the CL. We note that the
FM-FMD phase transition line is in a different universality class from 
the usual Higgs transition line, and is not continuously
connected to the symmetric phase that exists at small $\tk$ and
$\kappa$.  For a full account of the phase diagram, see ref.
\cite{BoGoSh97}.

We now introduce the 
fermion operators $\jn_{\rm R}=\j_{\rm R}$, $\jn_{\rm L}=\phi^{\dagger} \j_{\rm 
L}$, 
$\jc_{\rm L}=\j_{\rm L}$ and $\jc_{\rm R}=\phi \j_{\rm R}$. 
The fields with the superscripts c (charged)
and n (neutral) transform nontrivially under U(1)$_{\rm L}$ and 
U(1)$_{\rm R}$ respectively, both of which are unbroken only {\it at}
$\kappa=\kappa_c(\tk)$. 
We have calculated the neutral and charged fermion 
propagators to 1-loop order in perturbation theory in $1/\tk$ 
\cite{BoGoSh97b}. 
The fermion propagators can be written in the form  
\be
S^{{\rm n,c}}_{{\rm 1-loop}}(p)=\left[S^{-1}(p)  
+\S^{{\rm n,c}} (p) \right ]^{-1}\;,  \lb{PROP}
\ee
where $S^{-1}(p)= \sum_\m \{ i\g_\m \sin p_\m +2r \sin^2 \frac{p_\m}{2} \}$
is the inverse tree level (free fermion) propagator, and 
$\S^{{\rm n,c}} (p)$ is the 1-loop self-energy. A detailed discussion of 
$\S^{{\rm n,c}} (p)$ can be found 
in ref. \cite{BoGoSh97b}. Here we list the
crucial properties of the self-energy. 1) $\S^{{\rm n,c}} (0)=0$, implying
that no fermion-mass counterterm is needed
(consistent with
shift symmetry \cite{GoPe89}). 2) 
The doublers decouple because of the Wilson term in eq.~(\eq{PROP}).
$\S^{{\rm n,c}} (p)$ is regular at the 15
corners of the Brillouin zone and 
small compared to tree-level.
3) The RH (LH) component of $\S^{\rm c}(p)$ ($\S^{\rm n}(p)$)  
is a nonanalytic function of $p$ in the CL. For instance, 
for $\Sigma^c(p)$ we find in the limit $\k \searrow \k_c$ and at small $p$,
\be
 \S^c(p) \approx -i (32\p^2\tk)^{-1}\, \pslash\, P_{\rm R}\, \log(p^2) 
 \;, \lb{SI}
\ee
up to contact terms. Hence, nonanalytic  terms 
exist in the RH (but not in the LH) charged channel, and 
$S^{{\rm c}}_{{\rm R}x{\rm R}y}=\lag \jc_{{\rm R}x} \jbc_{{\rm R}y} \rag$
does not have an isolated pole.
In fact, 
$S^{{\rm c}}_{{\rm R}x{\rm R}y} = 
\lag \phi_x \jn_{{\rm R}x} \jbn_{{\rm R}y} \phi^{\dagger}_y \rag$ 
factorizes as   
$S^{{\rm n}}_{{\rm R}x{\rm R}y} \lag \phi_x \phi^{\dagger}_y \rag$ 
for large $|x-y|$.
$S^{{\rm n}}_{{\rm L}x{\rm L}y}$ factorizes in a similar manner, 
indicating the absence of $\phi$-$\jn_{\rm R}$
and $\phi^\dagger$-$\jc_{\rm L}$ bound states
(in contrast with the Smit-Swift model). 
Perturbation theory thus 
gives strong evidence that the spectrum contains  
only  LH charged and RH neutral fermions at $\k=\k_c(\tk)$.

%
\begin{figure}
\vspace*{-0.9cm}  
\begin{tabular}{c}
\epsfxsize=9.30cm
\hspace{-1.0cm} \epsfbox{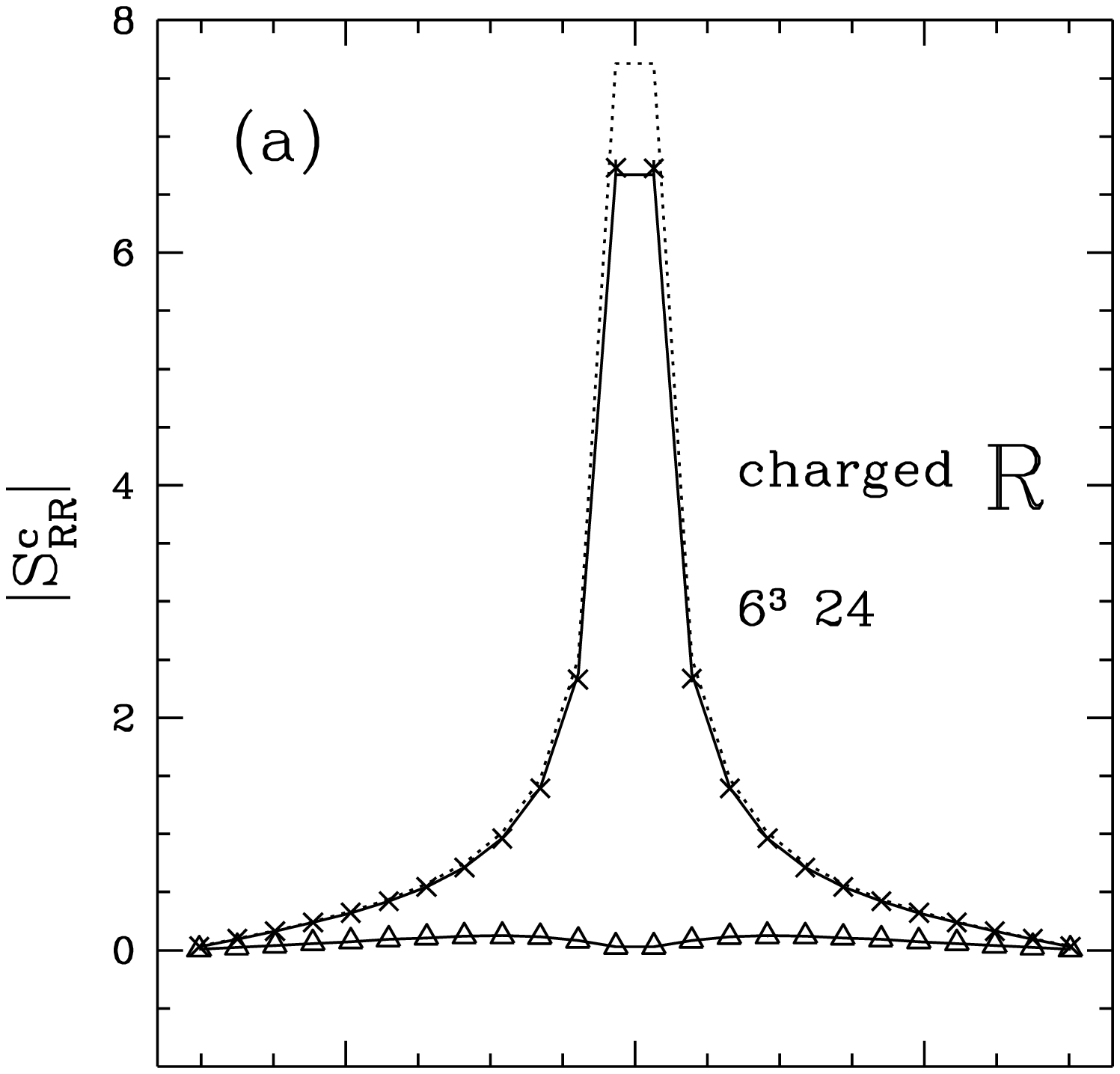}
\vspace*{-2.8cm} \\
\epsfxsize=9.30cm
\hspace{-1.0cm} \epsfbox{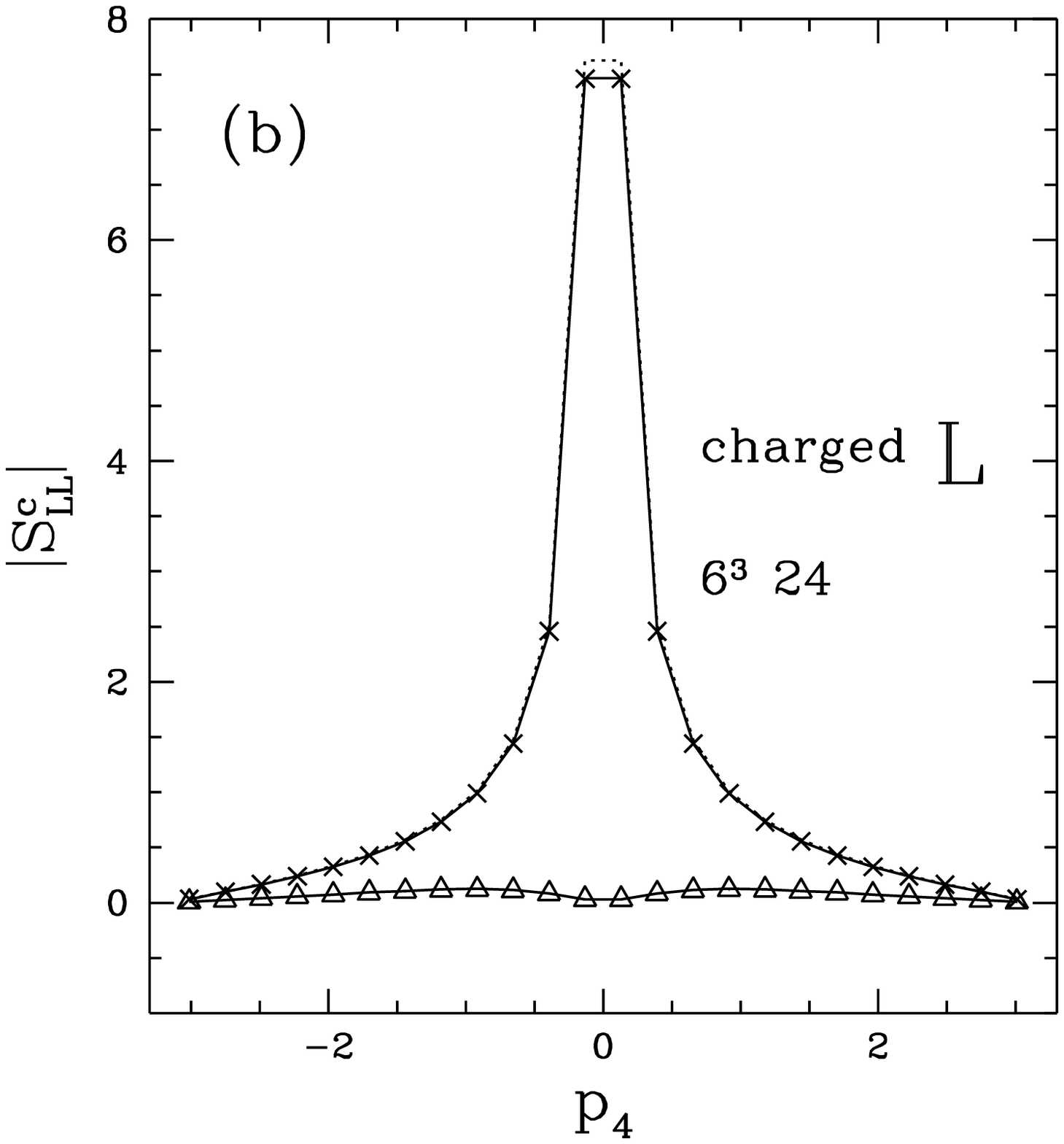}
\end{tabular}
\vspace*{-1.4cm} \\
\caption{The modulus of the charged fermion propagator 
components (a) $S^{{\rm c}}_{{\rm 
RR}}$  
and (b) $S^{{\rm c}}_{{\rm LL}}$ as a function of $p_4$. }
\label{FIG1}
\end{figure}
%
It is important to check that this result remains valid beyond 1-loop. 
To this end, for 
two momentum choices, $p=(0,0,0,p_4)$ and $p=(\p ,0,0,p_4)$, 
which allow us to probe the small momentum region as well as 
the edge of the Brillouin zone, we have computed 
the neutral and charged fermion propagators  
$S^{{\rm n,c}} (p) = -i[ S_{{\rm LL}}^{{\rm n,c}} (p) P_{\rm L} 
+S_{{\rm RR}}^{{\rm n,c}} (p) P_{\rm R}] \g_4 + S_{{\rm LR}}^{{\rm n,c}} (p)$, 
in the quenched approximation. 
All numerical computations were carried out at $\tk=0.2$ and $r=1$ 
on a $6^3 \times 24$ lattice.
The scalar field configurations were generated 
with a 5-hit Metropolis algorithm and 4000 configurations were skipped 
between fermionic measurements. We inverted the fermion matrix 
on 50 scalar field configurations. 
We used antiperiodic (periodic) boundary conditions in the
temporal (spatial) directions for the fermion fields, and 
periodic boundary conditions in all directions for the scalar field. 

To demonstrate that the species doublers decouple, we show 
in Fig.~\ref{FIG1} the modulus of the charged fermion 
propagator components  (a) $S_{{\rm RR}}^{{\rm c}} (p)$ and 
(b) $S_{{\rm LL}}^{{\rm c}} (p)$ as a function of $p_4$ 
for $p=(0,0,0,p_4)$ (crosses) and 
$p=(\p ,0,0,p_4)$ (triangles). We chose 
$\k=0.05$ which is 
in the FM-phase, very close to the FM-FMD phase 
transition \cite{BoGoSh97}.
The graphs show that there is no other pole in the Brillouin zone  
besides the one at the origin, 
and thus that the species doublers decouple. 
At first glance, Fig.~\ref{FIG1} suggests that 
both propagators have a pole at  $p=0$.
We will argue below that this is only true for
the LH case. 
The dotted and solid lines 
represent tree-level and 1-loop results,
obtained by 
evaluating the perturbative formula in eq.~(\eq{PROP})
on the same lattice, at the same point in the phase diagram. 
The                     
numerical data agree well with the 1-loop curve. The tree-level 
result agrees well with the numerical data only in 
Fig.~\ref{FIG1}b, indicating that $S_{{\rm LL}}^{{\rm c}}$
indeed behaves like a free propagator at small momenta. 
Similar graphs were obtained for the neutral propagators. 

\begin{figure}
\vspace*{-0.9cm} 
\begin{tabular}{c}
\epsfxsize=9.30cm
\hspace{-1.0cm} \epsfbox{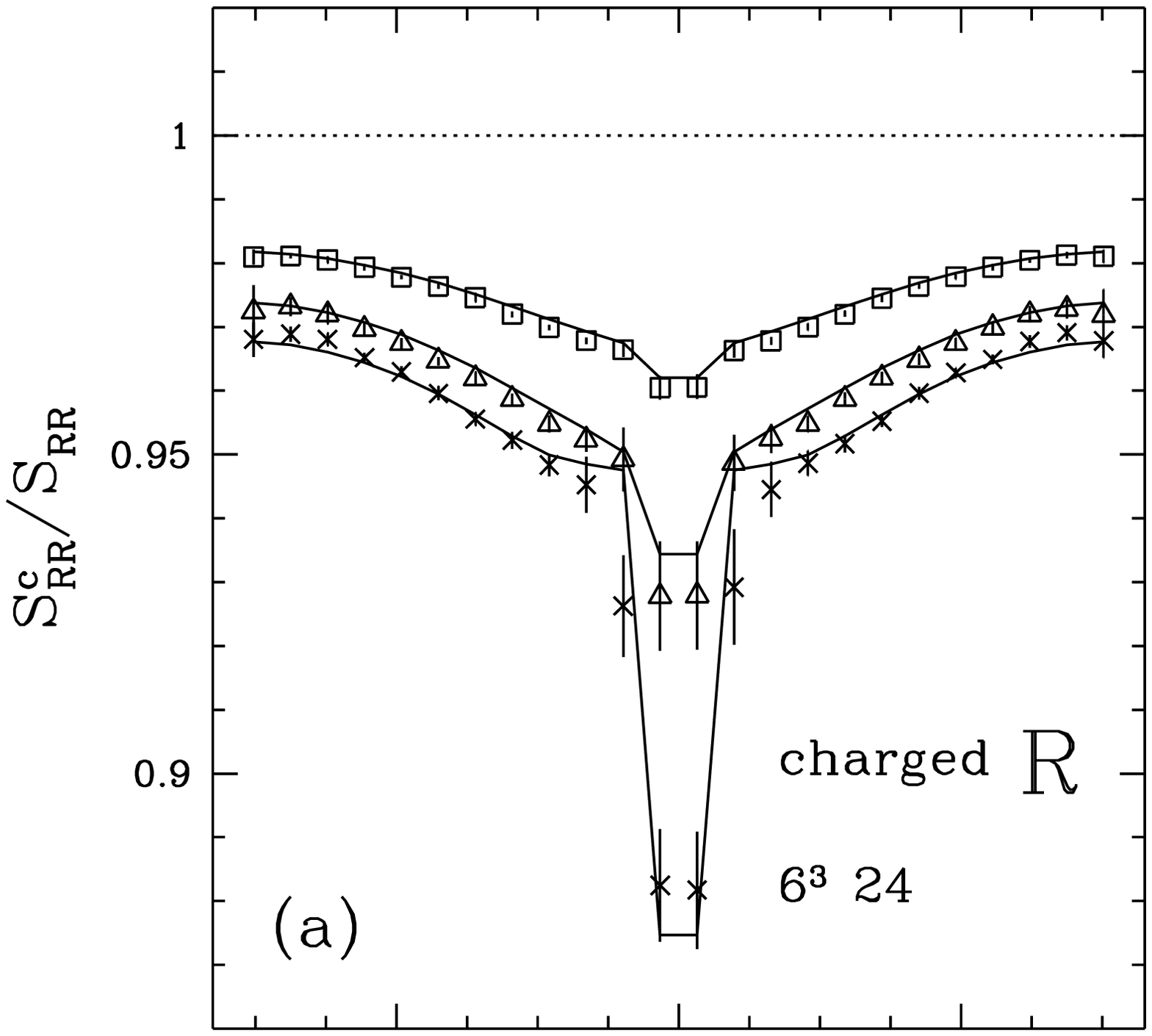}
\vspace*{-2.8cm} \\
\epsfxsize=9.30cm
\hspace{-1.0cm} \epsfbox{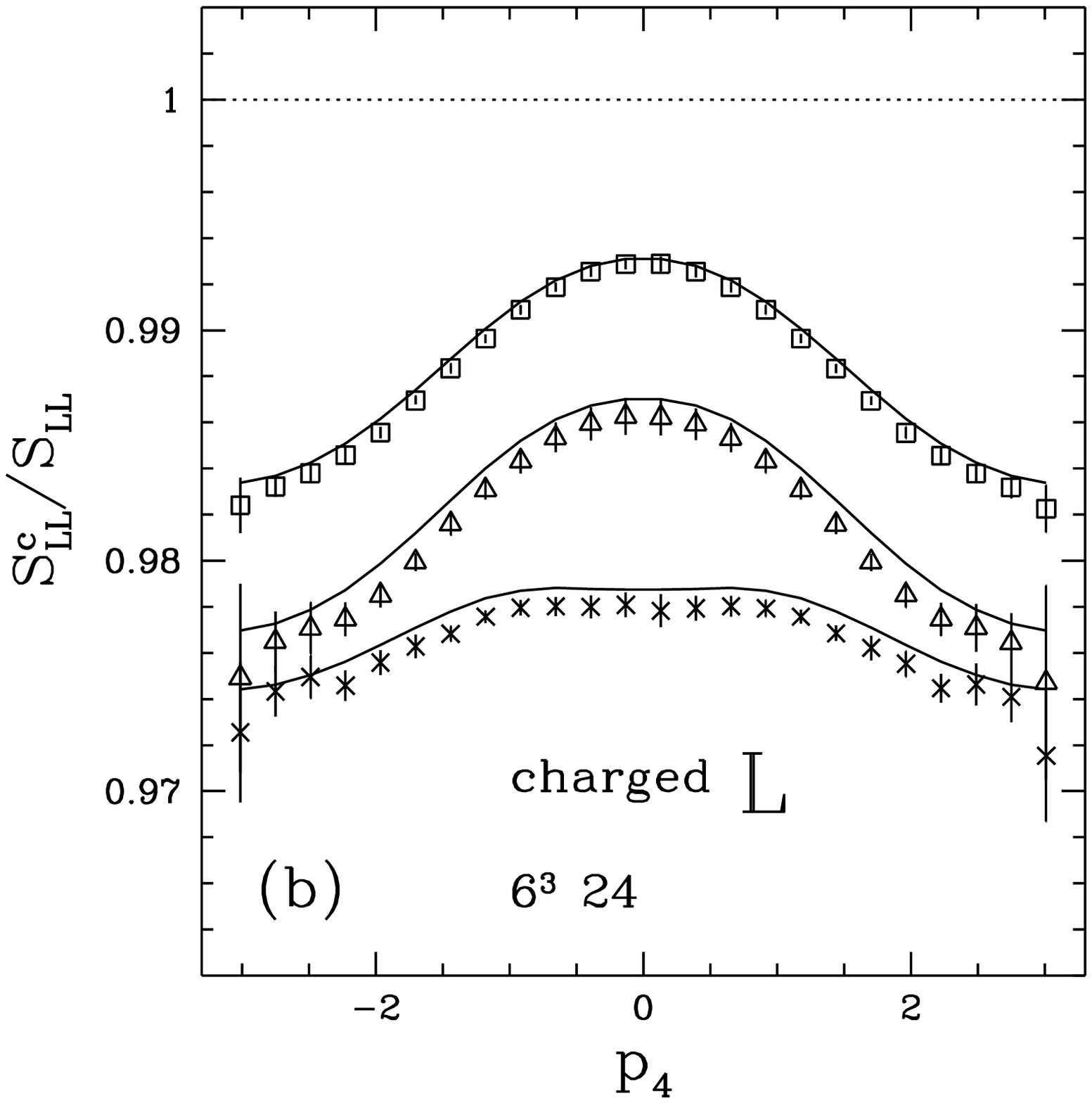}
\end{tabular}
\vspace*{-1.4cm} \\
\caption{The ratios (a) $S_{{\rm RR}}^{{\rm c}} /S_{{\rm RR}}$  
and (b) $S_{{\rm LL}}^{{\rm c}} /S_{{\rm LL}}$ as function of 
$p_4$.}
\label{FIG2}
\end{figure}

\begin{figure}
\vspace*{-0.9cm} 
\begin{tabular}{c}
\epsfxsize=9.30cm
\hspace{-1.0cm} \epsfbox{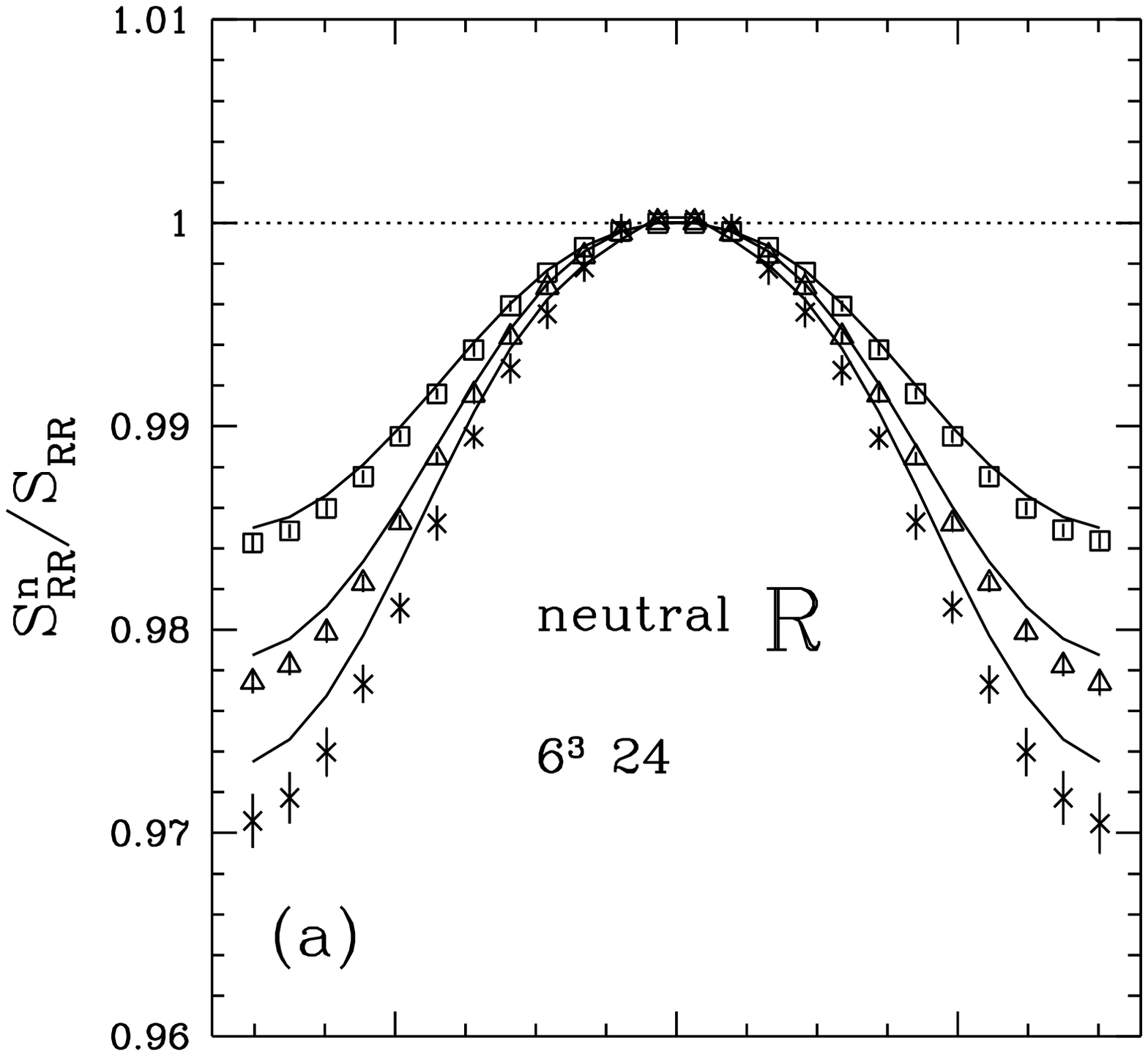}
\vspace*{-2.8cm} \\
\epsfxsize=9.30cm
\hspace{-1.0cm} \epsfbox{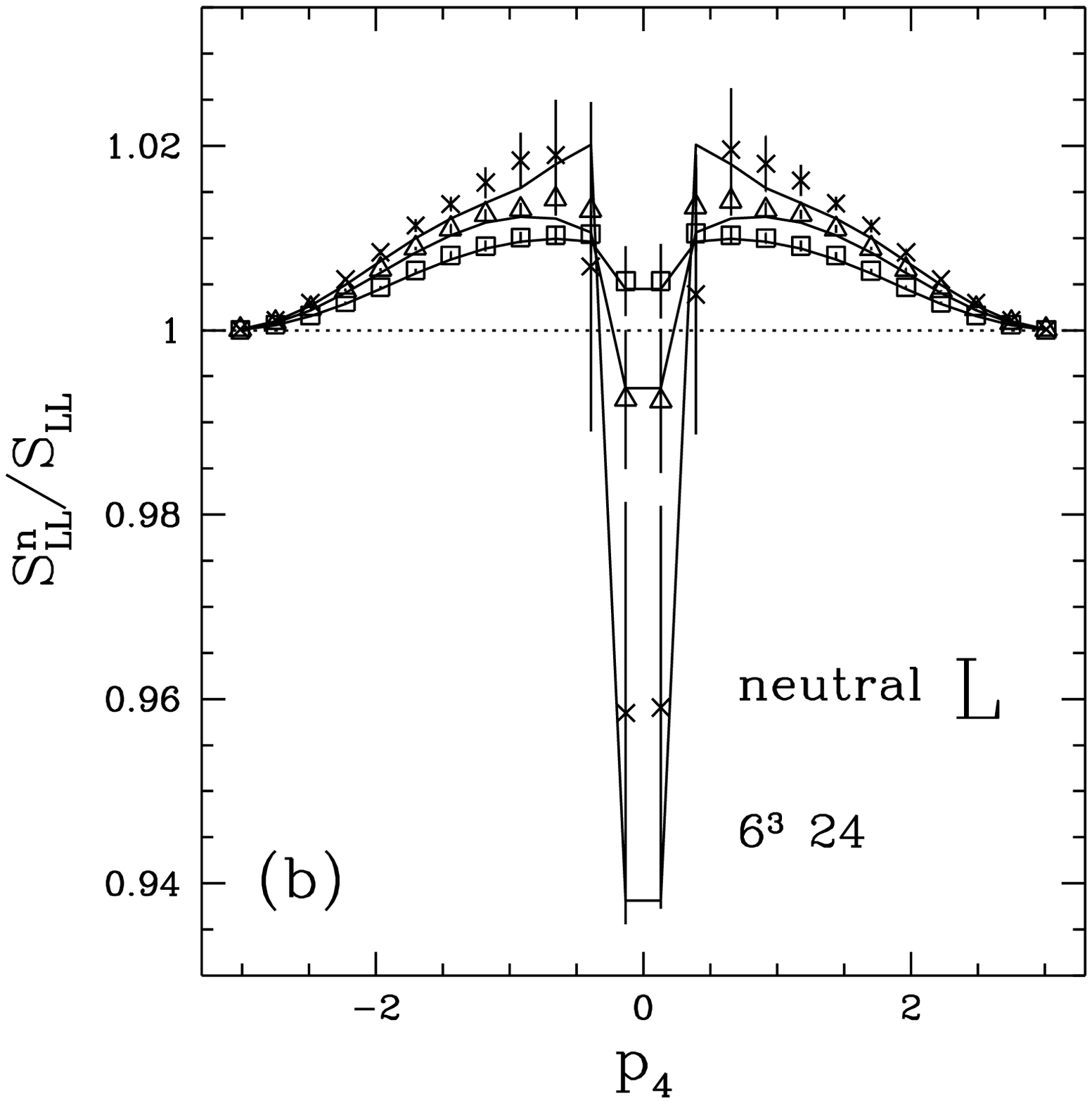}
\end{tabular}
\vspace*{-1.4cm} \\
\caption{The ratios (a) $S_{{\rm RR}}^{{\rm n}} /S_{{\rm RR}}$  
and (b) $S_{{\rm LL}}^{{\rm n}} /S_{{\rm LL}}$ as function of 
$p_4$.}
\label{FIG3}
\end{figure}

To figure out which of the four 
propagators has a pole 
at $p=0$ we have plotted in Fig.~\ref{FIG2}
the ratios $S_{{\rm RR}}^{{\rm c}}(p) /S_{{\rm RR}}(p)$
and $S_{{\rm LL}}^{{\rm c}}(p) /S_{{\rm LL}}(p)$,  
and in Fig.~\ref{FIG3} the ratios 
$S_{{\rm RR}}^{{\rm n}}(p) /S_{{\rm RR}}(p)$ and 
$S_{{\rm LL}}^{{\rm n}}(p) /S_{{\rm LL}}(p)$
for $p=(0,0,0,p_4)$ as a function of $p_4$. 
The ratios  marked in Figs.~\ref{FIG2} and \ref{FIG3} 
by crosses, triangles, and squares were obtained at 
$\k=0.05$, $0.3$ and $1$, which all fall in the FM phase \cite{BoGoSh97}. 
For $p_4 \ra 0$, the ratios should approach a line parallel to the abscissa, 
if the corresponding fermion propagator has a pole. 
Figs.~\ref{FIG2}b and \ref{FIG3}a indeed have a tangent with vanishing 
slope at $p_4=0$ for all values of $\k$, 
and hence $S_{{\rm LL}}^{{\rm c}}$ and $S_{{\rm RR}}^{{\rm n}}$ 
have a  pole at $p=0$.
The wave-function renormalization can be read off from the intercept 
at $p_4=0$. It is equal to one for $\jn_{\rm R}$ 
(because of shift symmetry \cite{GoPe89}) and smaller than 
one for $\jc_{\rm L}$.          
The ratios in Figs.~\ref{FIG2}a and \ref{FIG3}b exhibit 
a very different behavior. The cut in eq.~(\eq{SI}) manifests itself 
in the dip at small momenta, which becomes deeper 
for $\k\searrow\k_c$.                         
Dotted and solid lines 
denote again tree-level and 1-loop results, and
show that the data for the propagator 
ratios are in good agreement with 
the 1-loop curves. 

This agreement between perturbation theory and data in the FM phase
(with the small difference likely due to higher orders)
makes us confident that perturbation theory can be used to extrapolate
to the CL at the FM-FMD transition, with the conclusion that the spectrum
contains LH charged and RH neutral massless fermions.
\section{Summary and Outlook}
\label{SEC4}
We have shown that the spectrum of the 
reduced model consists, in the CL, of free LH charged and RH neutral 
fermions, decoupled from the unphysical $\phi$-sector.
A key element is our use of a gauge-fixed lattice theory which
exhibits the continuous phase transition described in section 3.
It should be interesting to study the dependence of these results
on the choice of gauge fixing.
 
Our conclusions hold for any choice of fermion species,
consistent with the fact that the anomaly vanishes in
the absence of a transversal gauge field.
Only the LH charged
%
%
\noindent
fermions couple to the gauge field when it is turned on
again. In that case, the unphysical states will  
decouple in the CL only if we choose an anomaly-free spectrum.

There are at least three important directions for future research:
1) Study of the U(1) case with full dynamical gauge fields. This
requires the fermion representation to be anomaly free. Although it
is technically difficult, we expect no problems of principle.
2) The extension of the gauge-fixing approach to the nonabelian
case is not trivial, simply because it is not known whether the
BRST formulation of gauge theories can be defined consistently
beyond perturbation theory. 3) A more technical issue concerns the problem
of fermion number violation \cite{fv}, on which 
work is in progress.\\

\noindent {\em Acknowledgements}: 
WB is supported by the Deutsche 
Forschungsgemeinschaft under grant Wo 389/3-2, MG by 
the US Department of Energy as an Outstanding Junior Investigator,
and YS by the US-Israel Binational Science
%
%
\noindent
Foundation, and the Israel Academy of Science.

\end{document}